\documentclass[%
 reprint,
superscriptaddress,
 amsmath,amssymb,
 aps,
 pre,
]{revtex4-1}

\usepackage{graphicx}
\usepackage{dcolumn}
\usepackage{bm}
\usepackage{xcolor}
\newcommand{\cor}{}

\newcommand{\di}{\mathrm{d}}

\begin{document}


\title{ { Steric interactions between mobile ligands facilitate complete wrapping in passive endocytosis} }

\author{Lorenzo Di Michele}
  \email{ld389@cam.ac.uk}
 \affiliation{
 Biological and Soft Systems, Cavendish Laboratory, University of Cambridge, JJ Thomson Avenue, Cambridge CB3 0HE, United Kingdom
 }
\author{Pritam Kumar Jana}%
\affiliation{%
 Universit\'e Libre de Bruxelles (ULB), Interdisciplinary Center for Nonlinear Phenomena and Complex Systems, Campus Plaine, CP 231, Blvd.\ du Triomphe, B-1050 Brussels, Belgium
}
\author{Bortolo Matteo Mognetti}
 \email{bmognett@ulb.ac.be}
\affiliation{%
 Universit\'e Libre de Bruxelles (ULB), Interdisciplinary Center for Nonlinear Phenomena and Complex Systems, Campus Plaine, CP 231, Blvd.\ du Triomphe, B-1050 Brussels, Belgium
}

\date{\today}

\begin{abstract}
Receptor-mediated endocytosis is an ubiquitous process through which cells internalize biological or synthetic nanoscale objects, including viruses, unicellular parasites, and nanomedical vectors for drug or gene delivery. In passive endocytosis the cell plasma membrane wraps around the ``invader'' particle driven by ligand-receptor complexation. By means of theory and numerical simulations, here we demonstrate how particles decorated by freely diffusing and non-mutually-interacting (ideal) ligands are significantly more difficult to wrap than those where ligands are either immobile or interact sterically with each other. Our model rationalizes the relationship between uptake mechanism and structural details of the invader, such as ligand size, mobility and ligand/receptor affinity, providing a comprehensive picture of pathogen endocytosis and helping the rational design of efficient drug delivery vectors.
\end{abstract}

\pacs{Valid PACS appear here}
\maketitle

{
\section{Introduction}
}
The cell plasma membrane is a complex interface, optimized to regulate {\cor cargo} transport. Internalisation of particles {\cor up to} a few tens of nanometers, including viruses and drug-delivery vectors, typically occurs \emph{via} endocytosis. In this process, a  particle (invader) is first wrapped by the membrane, and then internalized within an endosome~\cite{BookMolecularBiology}. Endocytosis is mediated by the binding of ligands, decorating the particle, to membrane receptors. The process is ``active" if aided by dedicated signaling pathways, as in clathrin-dependent~\cite{McMahon2011} and caveolin-dependent endocytosis~\cite{Nabi2003}, {\cor or} ``passive'', if solely mediated by multivalent ligand-receptor interactions, without energy consumption.  Viruses are sometimes able to hijack active endocytosis pathways, but in other cases are passively uptaken~\cite{Mercer2010,Boulant2015}. Artificial vectors, including solid nanoparticles~\cite{Zhang2015}, liposomes~\cite{Bareford2007,Duzgunes1999,Deshpande2013} and polymerosomes~\cite{Akinc2013}, are often uptaken by passive endocytosis. 
A deep understanding on how the  structure of the invader { and the molecular details of ligand-receptor interactions} influence passive endocytosis is thus required to aid the design of synthetic vectors, and to clarify the still poorly understood uptake mechanisms of pathogens \cite{dasgupta2014membrane,Mercer2010,Permanyer:aa}.\\
The modeling of passive endocytosis has traditionally relied on phenomenological approaches, where the multivalent nature of the interactions has been neglected \cite{dasgupta2017nano,tzlil2004statistical,chaudhuri2011effect},  or considered only in the limit of irreversible ligand-receptor binding~\cite{gao2005mechanics,decuzzi2008receptor,richards2016target,richards2014mechanism}. 
Thermodynamic models for multivalent interactions have instead been developed in the context of cell-cell adhesion~\cite{bell1984cell,coombs2004equilibrium,Krobath2009}, membrane targeting~\cite{kiessling2,kitov2003nature,licata2008kinetic,francisco-pnas,angioletti2017exploiting}, synapse formation~\cite{chen2003adhesion,coombs2004equilibrium,qi2001synaptic,raychaudhuri2003effective}, and the self-assembly of synthetic ligand-functionalized particles \cite{melting-theory1,theoryreview,bachmann2016melting}. These studies have highlighted how multivalent interactions give rise to complex phenomena, which can only be captured with a bottom-up modeling approach, { or Molecular Dynamics simulations~\cite{Vacha2011,Vacha2015}.} Particularly rich is the phenomenology observed in the presence of mobile linkers, which can freely diffuse on the substrates, and thus accumulate within the adhesion regions {~\cite{parolini2014thermal,shimobayashi2015direct,Zhdanov2017}.}  Receptors on cell-membranes and ligands on functionalized liposomes fall within this category~\cite{Deshpande2013}, while ligands on solid nanoparticles or viruses are often anchored to fixed points~\cite{Zhang2015,Karlsson-Hedestam:2008aa}. However, steric interactions between mobile linkers limit their local concentration~\cite{Dupuy:2008aa,Wasilewski:2012aa}, affecting adhesion in ways still unaccountable by state-of-art models.
\\
In this { paper}, we present an analytical and numerical description of passive endocytosis that correctly accounts for the multivalent nature of the interactions in the relevant scenarios of fixed and mobile ideal ligands and, for the latter, considers the effects of excluded volume interactions.
We demonstrate how particles functionalized by fixed ligands are more easily wrapped by the membranes, while  those functionalized by mobile ligands are the most prone to incomplete wrapping, due to the recruitment of the linkers within the adhesion regions.  Excluded volume interactions limit mobile-ligand accumulation, facilitating complete wrapping. { Accumulation of ligands is hindered by non-ideal entropic contributions that we estimate for the first time using perturbation theories and Monte Carlo simulations.}
\\
{ The remainder of this paper is structured as follows. In Sec.~\ref{SecII} we introduce the model and the simulation strategy. In Sec.~\ref{SecIIIa} we present the numerical framework employed to calculate adhesion free-energies. In Sec.~\ref{SecIIIb} we use the results of Sec.~\ref{SecIIIa} to study passive endocytosis of spherical and cylindrical invaders. In Sec.~\ref{SecIIIc} we study specific systems that have been considered by recent literature and corroborate the key role plaid by ligand mobility and steric interactions. Finally, in Sec.~\ref{SecIV} we summarize our results. \\
}
\begin{figure}[ht!]
\includegraphics[width=8.6cm]{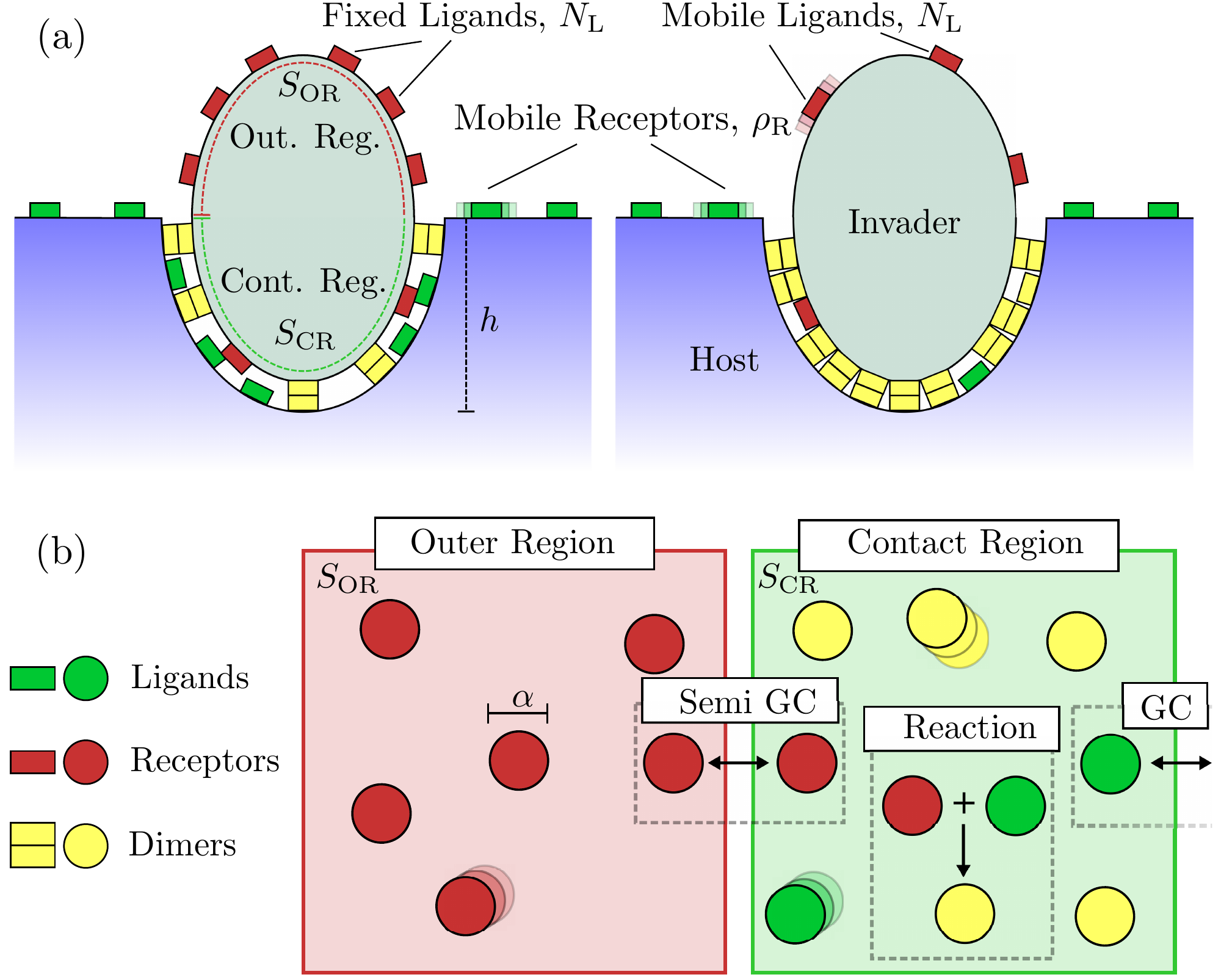} 
\caption{\label{Fig1} (color online) (a) Schematic of the invader-host interaction. The invader is decorated with either fixed  (left) or mobile (right) ligands, interacting with mobile receptors on the host surface. (b) Schematic of the Monte Carlo system used to compute adhesion free energy with non-ideal mobile ligands. Ligands, receptors and dimers are modeled as hard disks of diameter $\alpha$. The contact and outer regions are simulated as squares with periodic boundary conditions. Ligands are exchanged between the two regions \emph{via} semi-grand canonical moves, while receptors are exchanged with an ideal reservoir through grand canonical moves. Dimerization is controlled by a reaction move~\cite{Chen:2000aa}. See { App.~\ref{app:adhesion:MC} for the acceptance rules of all moves.}}
\end{figure}
{  \section{Modeling strategy}\label{SecII} }
We model the invader particle as a sphere or a prolate ellipsoid with axis of rotation orthogonal to the cell surface, penetrating to a depth $h$ (see Fig. \ref{Fig1}). The invader has total surface area $S_\mathrm{Tot}=S_\mathrm{CR}(h)+S_\mathrm{OR}(h)$, where $S_\mathrm{CR}(h)$ and $S_\mathrm{OR}(h)$ indicate the region in contact with the host cell and the non-adhering (outer) region, respectively. 
The overall interaction free energy between host and invader can be written as \cite{dasgupta2014membrane,lipowsky1992budding}
\begin{equation}\label{eqn:F}
F(h) = F_\mathrm{bend}(h) + F_\mathrm{stretch}(h)+F_\mathrm{lt}(h) + F_\mathrm{adh}(h).
\end{equation}
In Eq.~\ref{eqn:F}, $F_\mathrm{bend}(h) =  2 \kappa \int_{S_\mathrm{CR}(h)} \mathrm{d} S  \cdot  H^2$ is the membrane-bending contribution, where $\kappa$ is the bending modulus of the bilayer and $H$ is the mean curvature of the invader; $F_\mathrm{stretch}(h) = \sigma S_\mathrm{CR}(h)$ is the membrane-stretching contribution, with $\sigma$ indicating the stretching modulus; $F_\mathrm{lt} = \gamma \ell(h)$ accounts for all the line-tension effects, with $\ell(h)$ indicating the length of the triple line and $\gamma$ the line tension. { The expressions of $F_\mathrm{bend}$,  $F_\mathrm{stretch}$, and $F_\mathrm{lt}$ are shown in App.~\ref{app:elastic}. }
\\
{ In Eq.~\ref{eqn:F} and App.~\ref{app:elastic}, we neglect non-local elastic deformations of the host \cite{deserno2003wrapping}, as such terms are system specific (as proven by the large number of studies that neglected them) and do not influence the outcomes of our study. We prove this claim in Sec.~\ref{SecIIIc} where we combine our results of $F_\mathrm{adh}$ (see Eq.~\ref{eqn:F}) with the elastic contributions calculated in Ref.~\cite{Dasgupta2014}, the latter also accounting for the deformation of the membrane not in direct contact with the invader. } \\
The term $F_\mathrm{adh}(h)$ describes the ligand-receptor mediated adhesion. Previous studies have relied on the phenomenological assumption  $F_\mathrm{adh}(h) \propto S_\mathrm{CR}(h)$, while here we propose a representation that fully accounts for the multivalent nature of the interactions.
Since the invader is typically much smaller than the host, we can model the contact region as a finite surface of area $S_\mathrm{CR}$ in contact with an infinite reservoir of ideal  receptors. We indicate with $\rho_\mathrm{R}$ the average receptor density on the host. If no ligand-receptor complexes (dimers) are formed, and assuming that receptors can freely diffuse, the contact region should have receptor density $\rho_\mathrm{R}$. In our model $\rho_\mathrm{R}$ is controlled by the density of the ideal reservoir  $\rho_\mathrm{R}^{(0)}$, and by the extent of steric interactions between receptors. 
{\cor The assumption of ideal receptors produces $\rho_\mathrm{R} = \rho_\mathrm{R}^{(0)}$, while increasing steric repulsion causes $\rho_\mathrm{R}$ to decrease below $ \rho_\mathrm{R}^{(0)}$}. A number $N_\mathrm{L}$ of either fixed or mobile ligands is present on the invader.\\
The equilibrium constant $K^\mathrm{(eq)}_\mathrm{3D} = \exp(-\beta \Delta G_0) / \rho_\ominus$, where $\Delta G_0$ is the ligand-receptor interaction free-energy and $\rho_\ominus=1$M, controls dimerization in { 3D diluted solutions}. For linkers confined to a surface, a 2D equilibrium constant can be written as $K^\mathrm{(eq)}_\mathrm{2D} = \exp(-\beta \Delta G_0) / (\rho_\ominus \delta)$, where the $\delta$ is a length comparable with the size of the { linkers}, which accounts for entropic costs hindering dimerization~\cite{theoryreview,parolini2014thermal,shimobayashi2015direct}, and membrane roughness~\cite{xu2015binding,hu2013binding} { and deformability \cite{bachmann2016melting}}. 
{ For instance, when considering flexible rod--like linkers of length $L$ it can be shown that $\delta=L$ \cite{theoryreview,parolini2014thermal,shimobayashi2015direct}. For polymeric linkers, $K^\mathrm{(eq)}_\mathrm{2D}$ can be calculated using dedicated Monte Carlo algorithms \cite{bortolo-pnas,patrick-jcp,DeGernier2014}. } The specific form of $\delta$ does not affect { the results of this study,} so we describe dimerization propensity in terms of $K^\mathrm{(eq)}_\mathrm{2D}$.\\
Ligands, receptors and dimers are modeled as hard disks of diameter $\alpha$, which thus determines the extent of steric interactions { [see Fig.~\ref{Fig1}\,(b)]}. The same $\alpha$ is assumed for ligand-ligand, receptor-receptor, dimer-dimer, ligand-dimer and receptor-dimer interactions. Unbound ligands and receptors 
 {\cor are modeled as non-interacting.} The system { of Fig.~\ref{Fig1}} is then fully determined by the parameters $\{N_\mathrm{L}, \rho_\mathrm{R}^{(0)}, S_\mathrm{CR}, S_\mathrm{Tot}, K^\mathrm{(eq)}_\mathrm{2D}, \alpha\}$. \\

{ \section{Results}}

{ \subsection{Adhesion free energy}\label{SecIIIa}}

\begin{figure}
\includegraphics[width=8.6cm]{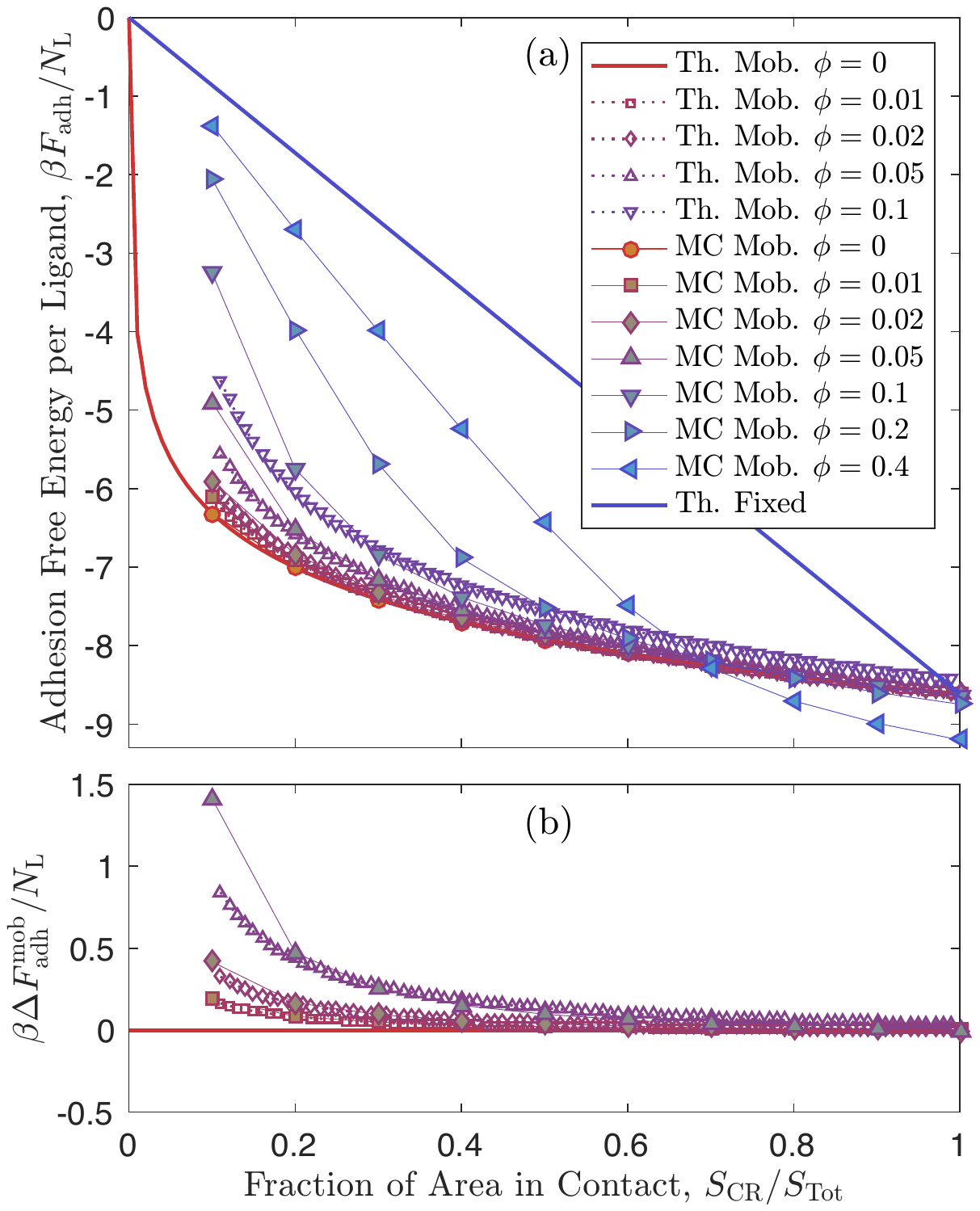} 
\caption{\label{Fig2} (color online) The influence of ligand mobility and steric interactions on the adhesion free energy. (a) Adhesion free energy as a function of the contact-area fraction calculated for fixed and mobile ideal ligands (solid lines, Eqs.~\ref{eq:fixed} and \ref{eq:ideal}) and non-ideal mobile ligands (dotted lines with empty symbols, Eq.~\ref{eq:excess}). Symbols connected by thin lines show the results of MC simulations. For the ideal cases we use $K^\mathrm{(eq)}_\mathrm{2D} \rho^{(0)}_R=5.5066\times10^3$. In the presence of steric interactions ($\phi>0$) we increase the reservoir receptor density $\rho^{(0)}_R$ to maintain a constant $\rho_R=N_\mathrm{L}/S_\mathrm{Tot}$. For $\phi=0.01, 0.02, 0.05, 0.1, 0.2, 0.4$ we scale $\rho^{(0)}_R$ (and thus $K^\mathrm{(eq)}_\mathrm{2D} \rho^{(0)}_\mathrm{R}$) by a factor $1.04, 1.09, 1.24, 1.57, 2.84, 20.5$, as calculated by dedicated MC simulations. In all cases we use $N_\mathrm{L} = 500$. (b) Deviation of the non-ideal adhesion free energy from the ideal case.}
\end{figure}

For $\alpha = 0$, the adhesive free energy $F_\mathrm{adh}(h)$ can be derived for both fixed and mobile ligands, analogously to what previously done in the context of linker-mediated particle interactions \cite{theoryreview,parolini2014thermal,shimobayashi2015direct}
\begin{eqnarray}
\beta F_\mathrm{adh}^{\mathrm{fix}, \alpha=0} (h) &=&  - N_\mathrm{L} \frac{S_\mathrm{CR}(h)}{S_\mathrm{Tot}} \log \left[1+K_\mathrm{2D}^\mathrm{(eq)}\rho_\mathrm{R}^\mathrm{(0)}\right] \label{eq:fixed}\\
\beta F_\mathrm{adh}^{\mathrm{mob}, \alpha=0} (h) &=&  - N_\mathrm{L} \log \left[1+\frac{S_\mathrm{CR}(h)}{S_\mathrm{Tot}} K_\mathrm{2D}^\mathrm{(eq)}\rho_\mathrm{R}^\mathrm{(0)}\right],\label{eq:ideal}
\end{eqnarray}
where in the ideal case $\rho^{(0)}_R=\rho_R$. { For completeness, in App.~\ref{app:adhesion:virial} and App.~\ref{app:adhesion:fixed} we report the explicit derivation, respectively, of Eq.~\ref{eq:ideal} and Eq.~\ref{eq:fixed} using exact evaluations of the partition function of the system. }
{ Eqs.~\ref{eq:fixed}, \ref{eq:ideal}} correctly account for all entropic contributions specific to multivalent interactions. For ideal fixed ligands, Eq.\,\ref{eq:fixed} recovers the phenomenological assumption $F_\mathrm{adh}(h) \propto S_\mathrm{CR}(h)$, with the advantage that our expression allows { linking} the proportionality constant to the microscopic details of the system. The trend determined in Eq.~\ref{eq:ideal} for the case of mobile ligands is instead {\cor strikingly} different, as shown in Fig.~\ref{Fig2}\,(a). The logarithmic dependence on $S_\mathrm{CR}$  translates into a sharp onset of adhesion, with the free energy flattening out as more of the invader gets wrapped{, as intuitively expected from the recruitment of ligands in the contact region.\\} 
\begin{figure*}
\includegraphics[width=15cm]{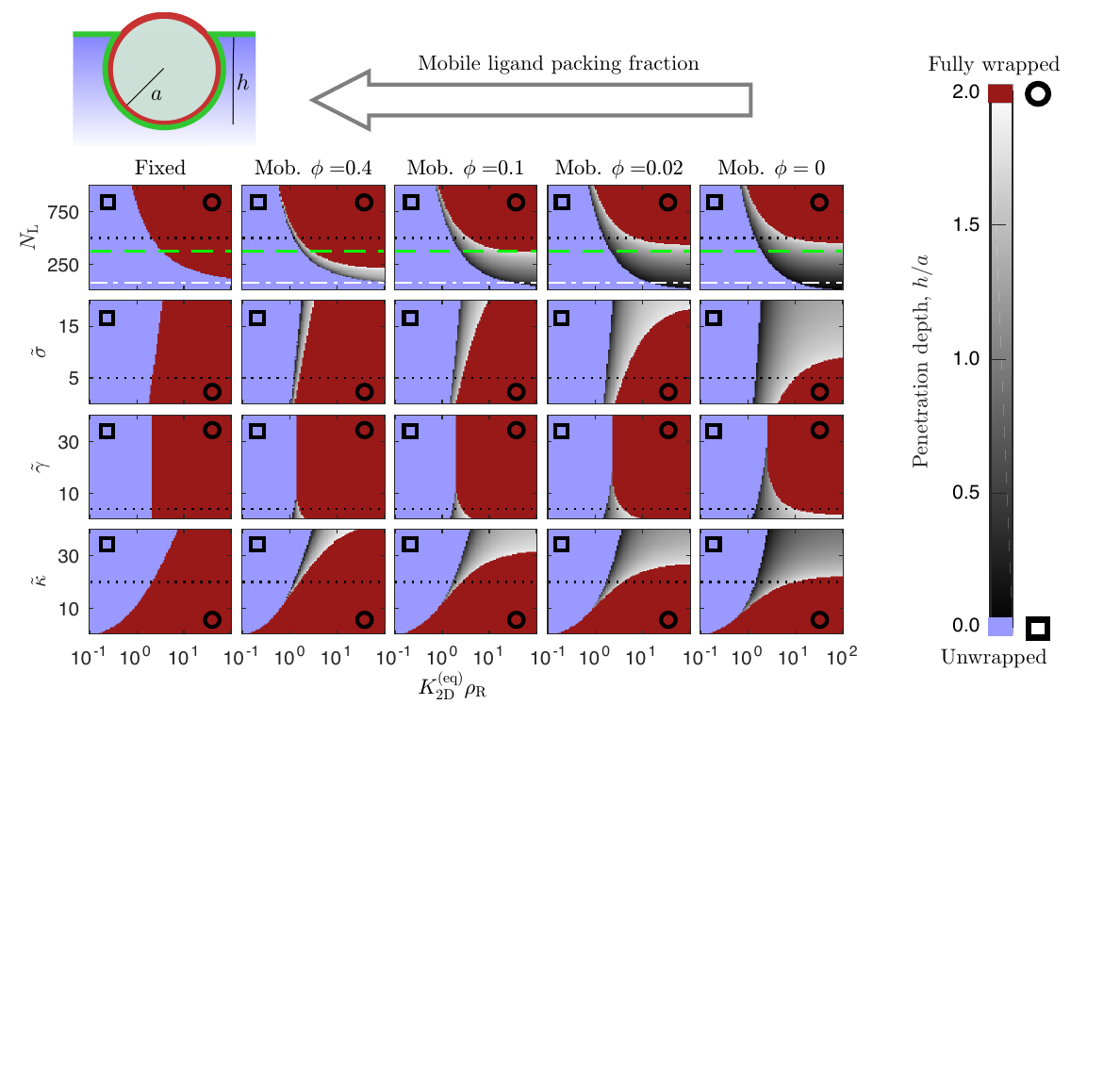} 
\caption{\label{Fig3} (color online) Mobile, thin ligands suppress complete wrapping of invader particles. The colormaps show the penetration depth $h\in[0,2a]$ { for spherical invaders of radius $a$} as a function of the { dimerization constant $K_\mathrm{2D}^\mathrm{(eq)}$ (in units of $\rho_R^{-1}$), } and the system parameters $N_L$, $\tilde{\kappa}=\kappa / [k_\mathrm{B}T]$, $\tilde{\gamma} = \gamma / [k_\mathrm{B}T a^{-1}]$, $\tilde{\sigma} = \sigma / [k_\mathrm{B}T a^{-2}]$. Regions corresponding to fully wrapped and unwrapped particles are marked by a circle and a square respectively. When not varied, the system parameters are fixed to the values marked by the dotted lines. { As for Fig.~\ref{Fig2}, we keep a constant $\rho_R=N_L/S_\mathrm{Tot}$ in all calculations.} For a typical invader size $a=50\,$nm and $T=37^\circ$C, these correspond to $\kappa = 8.6\times10^{-20}$ J, $\sigma = 8.5\times10^{-6}$ J\,m$^{-2}$, $\gamma = 3.4\times10^{-13}$ J\,m$^{-1}$, $N_\mathrm{L} = 500$. Green dashed lines and white dot-dash lines indicate respectively the number of ligands on typical influenza A (375) and HIV (73) virions~\cite{Harris:2006aa,Zhu:2006aa}.}
\end{figure*}
For ideal linkers, the accumulation of ligands is limited uniquely by { entropic costs ($\Delta S_\mathrm{cnf} \sim \log ( S_\mathrm{CR}/S_\mathrm{Tot})$)}, but for $\alpha>0$ steric interactions { further hinder ligand recruitment}. For non-ideal linkers the adhesive free energy can be written as $\beta F_\mathrm{adh}^{\mathrm{mob,} \alpha>0} (h) = \beta F_\mathrm{adh}^{\mathrm{mob} ,\alpha=0} (h) + \beta F_\mathrm{adh}^\mathrm{mob,ex} (h)$. The excess  free energy $\beta F_\mathrm{adh}^\mathrm{mob,ex} (h)$ can be evaluated through a second-order virial expansion ({ see App.~\ref{app:adhesion:virial}}) 
\begin{eqnarray}\label{eq:excess}
&&\beta F_\mathrm{adh}^\mathrm{mob,ex} (h) = N_\mathrm{L} B_2 K_\mathrm{2D}^\mathrm{(eq)} \left(\rho_\mathrm{R}^\mathrm{(0)}\right)^2 S_\mathrm{CR}  \times \\
&&~~~~\times {N_\mathrm{L}  K_\mathrm{2D}^\mathrm{(eq)} S_\mathrm{OR}/S_\mathrm{Tot} + 2  S_\mathrm{Tot} + 2 S_\mathrm{CR} K_\mathrm{2D}^\mathrm{(eq)} \rho_\mathrm{R}^\mathrm{(0)} \over \left(S_\mathrm{Tot} + S_\mathrm{CR}  K_\mathrm{2D}^\mathrm{(eq)}\rho_\mathrm{R}^\mathrm{(0)} \right)^2 },
\nonumber
\end{eqnarray}
 where $B_2 = \pi \alpha^2 / 2$ is the second virial coefficient of hard disks of diameter $\alpha$. In Fig.~\ref{Fig2} we demonstrate the effect of excluded volume, quantified by the ligand packing fraction $\phi = \pi \alpha^2 N_\mathrm{L} / S_\mathrm{Tot}$. As $\phi$ increases, the sharp adhesion onset as a function of $S_\mathrm{CR}$ becomes less evident. 
The analytical expansion in Eq.~\ref{eq:excess} is only accurate in the limit of small packing fraction. To access $F_\mathrm{adh}^{\mathrm{mob}, \alpha>0}$ at higher $\phi$, we adopt a Monte Carlo approach based on the model sketched in Fig.~\ref{Fig1}(b). The adhesion free energy is determined by thermodynamic integration~\cite{miriam}
\begin{equation}\label{Integration}
\beta F_\mathrm{adh}^{\mathrm{mob}, \alpha>0} = \int_{0}^{ K^\mathrm{(eq)}_\mathrm{2D} } \mathrm{d} K^\mathrm{(eq)}_\mathrm{2D} \frac{  \langle n_\mathrm{D} \rangle} {K^\mathrm{(eq)}_\mathrm{2D}},
\end{equation}
where $\langle n_\mathrm{D} \rangle$ is the average number of dimers estimated by MC at a given $K^\mathrm{(eq)}_\mathrm{2D}$.\\
The simulated adhesion free energy is shown in Fig.~\ref{Fig2}. For ideal linkers ($\phi=0$), we recover the result of Eq.~\ref{eq:ideal}, while for small { $\phi$} the numerical { and theoretical predictions match.} Deviations from the theory are observed at $\phi\gtrsim0.05$. When $\phi$ is further increased the adhesive free energy changes drastically, developing a linear region at low $S_\mathrm{CR}$ {analogous to the trend observed for fixed ligands.}
This behavior is a consequence of the excluded volume interactions frustrating the accumulation of ligands. 
Indeed, { even at large} $K^\mathrm{(eq)}_\mathrm{2D}$, the number of ligands to get recruited in the contact region { is limited by the diverging chemical potential of packed hard disks.} { Consequently, adhesion free--energies become linear at low $S_\mathrm{CR}$.} Surprisingly, at high $\phi$ (\emph{e.g.} $\phi=0.4$) and large $S_\mathrm{CR}$ we observe that $F_\mathrm{adh}^{\mathrm{mob}, \alpha>0}$ becomes more attractive than $F_\mathrm{adh}^{\mathrm{mob}, \alpha=0}$. This effect is caused by a reduction in the overall steric hindrance following dimerization: The area excluded to each dimer by an unbound ligand-receptor pair is larger than the area excluded by a single dimer.\\

\begin{figure*}[ht!]
\includegraphics[width=15cm]{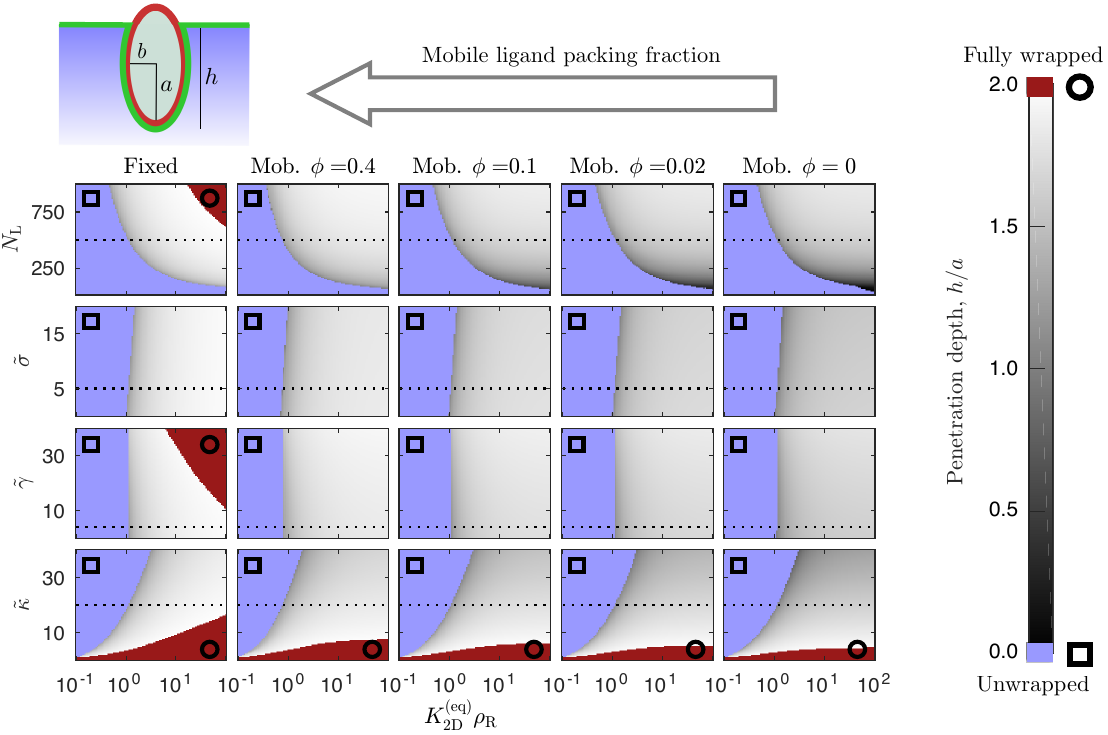} 
\caption{ 
(color online) Penetration depth for prolate invaders as a function of $K_\mathrm{2D}^\mathrm{(eq)}$ and the system parameters (see the caption of Fig.~\ref{Fig3} for definitions). The invader is taken with the semi-major axis $a$  perpendicular to the host surface and semi-minor axis equal to $b=a/2$. When not varied, the system parameters are fixed to the values marked by the dotted lines and, if $a=50\,$nm and $T=37^\circ$C, correspond to the values reported in the caption of Fig.~\ref{Fig3}. Regions corresponding to fully wrapped and unwrapped particles are marked by a circle and a square respectively.
}\label{FigureProlate}
\end{figure*}

 \subsection{Endocytosis phase--diagrams }\label{SecIIIb}
{ To study the effect of ligand mobility  and excluded volume interactions on endocytosis, we combine Eq.~\ref{eqn:F} with the analytical expressions for $F_\mathrm{adh}$ in the regimes of fixed and mobile ideal ligands (Eqs.~\ref{eq:fixed} and \ref{eq:ideal}), and with numerical estimates of $F_\mathrm{adh}^{\mathrm{mob}, \alpha>0}$ for sterically interacting mobile ligands (Eq.~\ref{Integration}).
We focus on invader of spherical shape, mimicking artificial nanoparticles, liposomes and many enveloped viruses including HIV and influenza~\cite{Karlsson-Hedestam:2008aa,Harris:2006aa}.}  The overall free energy is minimized as a function of the penetration depth $h \in [0, 2a]$, { where $a$ is the radius of the invader}, and the equilibrium values of $h$ are shown in Fig.~\ref{Fig3} as a function of { ${K^\mathrm{(eq)}_\mathrm{2D}}\rho_\mathrm{R}$} (cfn.\ Eqs.~\ref{eq:fixed} and \ref{eq:ideal}) and all the other relevant system parameters: $N_\mathrm{L}$, $\sigma$, $\kappa$, and $\gamma$. For generality, membrane tension, bending modulus, and line tension are expressed in reduced units $\tilde{\sigma} = \sigma / [k_\mathrm{B}T a^{-2}]$, $\tilde{\gamma} = \gamma / [k_\mathrm{B}T a^{-1}]$ and $\tilde{\kappa} = \kappa / [k_\mathrm{B}T]$. For temperature $T=37^\circ$C, and considering invader similar in size to a typical virus, \emph{i.e.} $a=50$\,nm, the range of parameters covered in Fig.~\ref{Fig3} spans biologically relevant intervals $\kappa  \in [0,30] \times k_B T$~\cite{Yoon:aa,Popescu:2006aa,Scheffer:2001aa}, $\sigma \in [0,34]\times10^{-6}$ J\,m$^{-2}$~\cite{Yoon:aa,Popescu:2006aa,Lieber:2013aa}, $\gamma \in [0,12]\times10^{-13}$ J\,m$^{-1}$~\cite{dasgupta2014membrane,Garcia-Saez:2007aa}, and $N_\mathrm{L} \in [10,1000]$~\cite{Harris:2006aa,Zhu:2006aa}.\\
Spherical invaders featuring fixed ligands always display a first-order transition between fully unwrapped ($h=0$) and fully wrapped ($h=2a$) configurations. Partially wrapped states do not occur, as previously observed when neglecting long-range elastic deformations of the host membrane~\cite{deserno2003wrapping}.
As intuitively expected, the wrapping transition occurs at lower { $K^\mathrm{(eq)}_\mathrm{2D}$}  for ``softer" membranes (lower $\tilde{\sigma}$ and $\tilde{\kappa}$) and higher number of ligands on the invader. No $\gamma$-dependence is observed, since  { $\ell=0$} in both fully wrapped and fully unwrapped states.\\ 
The scenario changes drastically for the case of mobile ligands with {$\phi=0$}, were we observe the emergence of several partially wrapped configurations. The phase boundary marking the onset of wrapping differs only marginally from the case of fixed ligands, but the range of conditions where full wrapping is achieved is significantly reduced. For instance, while with fixed ligands and $K^\mathrm{(eq)}_\mathrm{2D}\rho^{(0)}_\mathrm{R} = 10^2$ full wrapping is reached at $N_\mathrm{L} \simeq 140$, for mobile ligands $N_\mathrm{L}$ needs to be as large as 540. Likewise, for the same ligand-receptor affinity, fixed ligands induce full wrapping at all tested values of $\tilde{\kappa}$ and $\tilde{\sigma}$, while for mobile ligands $\tilde{\kappa}<22$ and $\tilde{\sigma}<9$ are required, values that can easily be exceeded in typical biological cells~\cite{Popescu:2006aa,Lieber:2013aa}. The reduced tendency to complete wrapping is a direct consequence of the rearrangement of the mobile ligands, whose accumulation within small contact regions suppresses the enthalpic drive for further wrapping.  As expected from the trends shown in Fig.~\ref{Fig2} for the adhesion free energy in the presence of steric interactions, increasing ligand packing fraction in the mobile regime favours complete wrapping, recovering a behaviour not dissimilar from that of fixed ligands for $\phi=0.4$. \\
 { In Fig.~\ref{FigureProlate}} we assess the wrapping behaviour of an invader shaped like a prolate ellipsoid, resembling the malaria plasmodium~\cite{dasgupta2014membrane}. { The ellipsoid is arranged with the major axis perpendicular to the surface of the host, mimicking the invasion geometry of the malaria plasmodium~\cite{dasgupta2014membrane}.} 
  { In this case,} partially wrapped states are present also for the case of fixed ligands. However, mobile ligands cause the regions of stable fully wrapped configurations to shrink significantly, and in some cases disappear altogether from the tested parameter range. Moreover, the partially wrapped states found with fixed ligands tend to be close to full wrapping, while mobile ligands tend to stabilize marginally wrapped configurations. { As for spherical invaders (Fig.~\ref{Fig3}), steric interactions between ligands facilitate wrapping.}\\
Calculations of Figs.~\ref{Fig3}, \ref{FigureProlate} neglect the energetic terms related to the deformation of the non-adhering part of the host membrane. { Therefore in the next section we calculate the degree of wrapping} using previously published numerical estimates of the membrane-deformation energy, fully accounting for non-local deformations~\cite{Dasgupta2014}. 
\\

\begin{figure}[ht!]
\includegraphics[width=8.6cm]{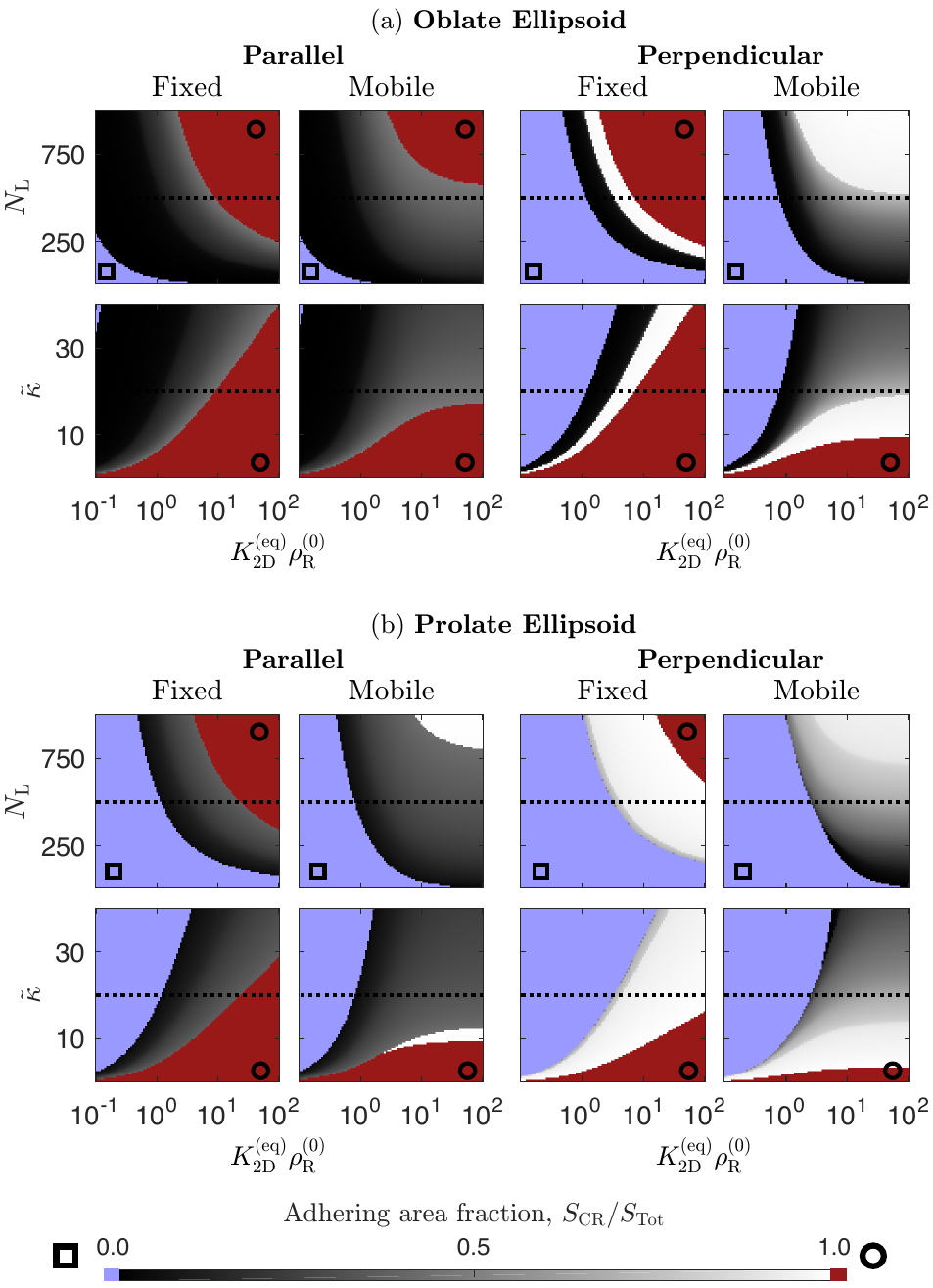} 
\caption{ (color online) Equilibrium wrapping degree of ellipsoidal particles at varying number of ligands $N_\mathrm{L}$ and bending rigidity $\tilde{\kappa}$. The shape of the invader is described by $(x^2 + y^2)/a^2 + (|z|/b)^2 = 1$ with $b/a=0.8$ (top) and $b/a=1.75$ (bottom). Please refer to the caption of Fig.~\ref{Fig3} for the definition of the adimensional system parameters. When not varied, the bending rigidity has been taken equal to $\tilde \kappa=20$, and the number of ligands to $N_\mathrm{L} = 500$ (dotted lines). The membrane tension is set to $\tilde \sigma = 16$ by Dasgupta \emph{et al.}~\cite{Dasgupta2014}. Regions corresponding to fully wrapped and unwrapped particles are marked by a circle and a square respectively. 
}\label{fig:S1b}
\end{figure}

\begin{figure}[ht!]
\includegraphics[width=8.6cm]{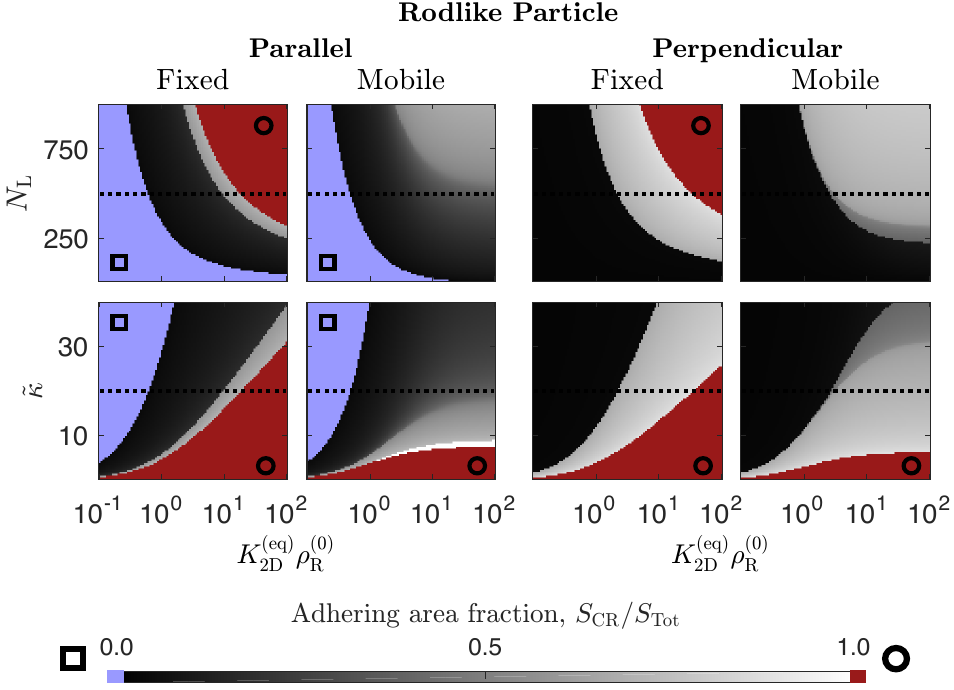} 
\caption{ (color online) Equilibrium wrapping degree of rod-like particles at varying number of ligands $N_\mathrm{L}$ and bending rigidity $\tilde{\kappa}$. The shape of the invader is described by the equation  $[(x^2 + y^2)/a^2]^{(n/2)} + (|z|/b)^n = 1$ with $n=4$ and $b/a=1.5$. Please refer to the caption of Fig.~\ref{Fig3} for the definition of the adimensional system parameters.  When not varied, the bending rigidity has been taken equal to $\tilde \kappa=20$, and the number of ligands to $N_\mathrm{L} = 500$ (dotted lines); The membrane tension is set to $\tilde \sigma = 16$ by Dasgupta \emph{et al.}~\cite{Dasgupta2014}. Regions corresponding to fully wrapped and unwrapped particles are marked by a circle and a square respectively.
}\label{fig:S1c}
\end{figure}

\subsection{Effect of non-adhering membrane and nanoparticle orientation}\label{SecIIIc}

In this section we study the wrapping behavior using a free energy functional $F$ in which the membrane-deformation energy terms are replaced by the numerical estimate $F_\mathrm{elast}$, extracted from Ref.~\cite{Dasgupta2014}
\begin{equation}\label{eq:num}
F(S_\mathrm{CR}/S_\mathrm{Tot}) = F_\mathrm{elast}(S_\mathrm{CR}/S_\mathrm{Tot}) + F_\mathrm{adh}(S_\mathrm{CR}/S_\mathrm{Tot}).
\end{equation}
In Eq.~\ref{eq:num}, $F_\mathrm{elast}$ accounts for the membrane bending, stretching, and the tension of the host-invader contact line (respectively $F_\mathrm{bend}$, $F_\mathrm{stretch}$ and $F_\mathrm{lt}$ in Eq.~\ref{eqn:F}), but also for the non-local elastic contribution of the deformed portion of the membrane surrounding the invader. For detailed information on the calculation of $F_\mathrm{elast}$ and the chosen boundary conditions we refer to Ref.\ \cite{Dasgupta2014}. 
Note that in Eq.~\ref{eq:num}, the free energy is expressed as a function of the fraction of the invader area in contact with the host $S_\mathrm{CR}/S_\mathrm{Tot}$, rather that the penetration depth $h$. $F_\mathrm{elast}$ has been extracted by fitting digitalized data of Ref.\ \cite{Dasgupta2014} (see Fig. S1, supporting materials of Ref.~\cite{Dasgupta2014}) using polynomials of degree ten, while $F_\mathrm{adh}$ is given by Eqs.~\ref{eq:fixed} and \ref{eq:ideal}, for fixed and mobile ligands respectively. The equilibrium wrapping state is determined by numerically minimizing $F$ as function of $S_\mathrm{CR}/S_\mathrm{Tot}$.
In Figs.~\ref{fig:S1b} and \ref{fig:S1c} we show the equilibrium fraction of contact area for invaders shaped like prolate/oblate ellipsoids and rods, oriented with their symmetry axis parallel or perpendicular to the host surface. The degree of wrapping is mapped as a function of $K_\mathrm{2D}^\mathrm{eq} \rho_\mathrm{R}^{\mathrm(0)}$ and either $N_\mathrm{L}$  or $\tilde\kappa$. In all cases, we observe the same qualitative trends shown in Fig. \ref{Fig3} and \ref{FigureProlate}, and calculated using the analytical expression for the interaction free energy  (Eq.~\ref{eqn:F}).  This demonstrates how the effect of ligand mobility on the wrapping behavior is largely insensitive on the details of membrane mechanics. In particular, the range of parameters for which fully wrapped configurations are stabilized is strongly suppressed for the case of mobile ligands, which tend to induce partial wrapping due to the accumulation of linkers in the contact region. For fixed linkers, in turn, we observe a greater tendency towards complete or near-complete wrapping, caused by the uniform adhesion force and the resulting enthalpic drive for maximizing the contact area. Different invader shapes and orientations display different wrapping behaviors, as thoroughly discussed by Dasgupta \emph{et al.}~\cite{Dasgupta2014}. Note also the semi-quantitative agreement between the patterns calculated with the two methods for prolate-ellipsoidal invaders oriented perpendicular to the host surface (bottom right in Fig.~\ref{fig:S1b}  and right hand-side of Fig.~\ref{Fig3}). 
\\

\section{Conclusions}\label{SecIV}

In summary, we apply state of art modeling of ligand-mediated-interactions to the problem of passive endocytosis, and demonstrate how membrane wrapping of invader particles is drastically affected by ligand mobility and steric interactions. If ligands are diffusive and have negligible steric interactions, complete membrane wrapping is hindered, and the invading particle is often found in a partially engulfed state. In turn, complete membrane wrapping is facilitated if ligands are immobile or their accumulation is substantially limited by steric interactions.\\
These effects may have important implications in understanding the relationship between the structure of biological invaders and their ability to induce passive endocytosis, { which would be particularly relevant in the context of viral invasion, where several competing uptake pathways have been observed or hypothesized~\cite{Mercer2010,Boulant2015}}. Regardless of the capsid shape, many viruses are enveloped by a (near) spherical lipid bilayer, decorated with glycoprotein complexes (spikes), whose role is targeting cell receptors and driving endocytosis~\cite{Harris:2006aa}. Despite being embedded in a lipid membrane, these ligands are anchored to a protein matrix present underneath the bilayer, which makes them immobile~\cite{Harris:2006aa,Zhu:2006aa}. Our results suggest that ligand anchoring may be crucial to allow or at least facilitate membrane wrapping in enveloped viruses. Indeed, influenza A has $\sim375$ spikes on its surface~\cite{Harris:2006aa,Zhu:2006aa}, which according to our model may not be sufficient to induce complete wrapping if the ligands were mobile (Fig.~\ref{Fig3}). In turn we predict that, with only $\sim73$ ligands on its surface, HIV virions would struggle to achieve passive engulfment even in the regime of fixed ligands, suggesting that active endocytosis pathways may be a strict requirement~\cite{Permanyer:aa}.\\
Our findings apply as well to artificial delivery vectors relying on passive endocytosis. For instance, we predict that passive endocytosis of functionalized liposomes can be enhanced by choosing high-viscosity lipid formulations that hinder ligand mobility, or choosing high-molecular weight ligands to boost their packing fraction. { 
Future designs of delivery systems will likely need to refine the molecular properties of the ligands (see, e.g., Ref.~\cite{C7NR09011K}). In this respect, our manuscript provides a valuable design platform allowing to study the impact of molecular details of the ligands on the degree of wrapping. For instance, changes in the ligand--receptor affinities close to the triple line, as due to larger host-guest distances, could be easily investigated using heterogeneous association constants ($K^\mathrm{(eq)}_{2D}$). }
\\ 
\begin{acknowledgments}
The authors thank Pietro Cicuta for fruitful discussions and comments on the manuscript. The work of BMM and PKJ was supported by the Fonds de la Recherche Scientifique de Belgique - FNRS under grant n$^\circ$ MIS F.4534.17. LDM acknowledges support from Emmanuel College Cambridge, the Leverhulme Trust, and the Isaac Newton Trust through an Early Career Fellowship (ECF-2015-494), the Royal Society through a University Research Fellowship (UF160152) and the EPSRC Programme Grant CAPITALS number EP/J017566/1. Computational resources have been provided by the Consortium des \'Equipements de Calcul Intensif (CECI), funded by the Fonds de la Recherche Scientifique de Belgique - FNRS under grant n$^{\circ}$  2.5020.11.
\end{acknowledgments}

\appendix

{

\section{ Elastic deformation of the membrane}\label{app:elastic}

To analytically estimate the energy cost associated to the deformation of the membrane, used to calculate the wrapping phase diagrams in Figs.~\ref{Fig3} and  \ref{FigureProlate} (Eq.~\ref{eqn:F}), we model the invader as a prolate ellipsoid with axis (${\bf z}$) orthogonal to cell surface, defined by the equation
\begin{eqnarray}
{x^2 \over a^2} + {y^2 \over a^2} + {z^2 \over b^2} = 1,
\end{eqnarray}
with $b>a$ and eccentricity defined by
\begin{eqnarray}
e^2 &=& 1 - { a^2 \over b^2 }.
\end{eqnarray}
The case of spherical invader is simply recovered in the limit $a=b$ and $e=0$. In polar coordinates $\theta$ and $\varphi$, the surface of the invader is parametrised as 
\begin{eqnarray}
x = a \cos \theta \cos \varphi
\quad
y = a \cos \theta \sin \varphi 
\quad
z = b \sin \theta.
\end{eqnarray}
As detailed in Eq.~\ref{eqn:F}, the free energy of the system, in which the innermost point of the invader penetrates to a depth $h$,
comprises a membrane stretching term ($F_\mathrm{stretch}$), membrane bending term ($F_\mathrm{bend}$), a line tension term ($F_\mathrm{lt}$), and an adhesion term 
($F_\mathrm{adh}$). Below we calculate the energy terms associated with the mechanical deformation of the membrane \cite{helfrich1973elastic}.
\\

{\bf Membrane stretching.} { The stretching  energy is calculated as $F_\mathrm{stretch}=\sigma S_\mathrm{CR}(h)$, where $S_\mathrm{CR}(h)$ is the contact area between invader and cell and $\sigma$ is the cell-membrane stretching modulus. Defining $y(h) = (h-b)/b$, in the general case we find}
\begin{eqnarray}
F_\mathrm{stretch}^\mathrm{ellips}(h)
&=& 2\pi a b \sigma \int_{-1}^{y(h)} \mathrm{d} y \sqrt{1-e^2 y^2}
\nonumber \\
 &=& 2 \pi a b \sigma \Big[ {y(h) \over 2} \sqrt{1 - e^2 y(h)^2 } + { \arcsin (e y(h)) \over 2 e} 
 \nonumber \\
 && + {1\over 2}  \sqrt{1 - e^2 } +   { \arcsin e  \over 2 e}
\Big]
\\
{\mathrm{d} F_\mathrm{stretch}^\mathrm{ellips}(h) \over \mathrm{d} h} &=& 2 \pi a \sigma \sqrt{1-e^2  y(h)^2},
\nonumber
\end{eqnarray}
while for spherical invaders we obtain
\begin{eqnarray}
F_\mathrm{stretch}^\mathrm{sph}(h)
= 2\pi a h \sigma 
&\qquad & 
{\mathrm{d} F_\mathrm{stretch}^\mathrm{sph}(h) \over \mathrm{d} h} = 2 \pi a \sigma.
\end{eqnarray}

{\bf Membrane bending.} The bending energy of the membrane calculated as the integral over the contact area of $2 \kappa H^2$ where $\kappa$ is the bending modulus and $H$ is the average curvature $H = 1/( 2 c_1)  +  1 / (2 c_2)$, where $c_1$ and $c_2$ are the principal radii of curvature at a given point ${\bf r}=(x,y,z)$. These radii are equal to $\alpha^2/p$ and $\beta^2/p$ where $\alpha$ and $\beta$ are the semi-axes of the ellipse obtained intersecting the ellipsoid with the central plane parallel to the plane tangent to ${\bf r}$, while $p$ is the distance between the center of the ellipsoid and the tangent plane \cite{m1929principal}. We find 
\begin{eqnarray}
& \alpha^2 = a^2 \qquad
 \beta^2 = a^2 \sin^2 \theta + b^2 \cos^2 \theta &
 \nonumber \\
 & p = {a b \over \sqrt{a^2 \sin^2 \theta + b^2 \cos^2 \theta} }, &
\end{eqnarray}
 resulting in 
 \begin{eqnarray}
 H &=&  {a b \over \sqrt{a^2 \sin^2 \theta + b^2 \cos^2 \theta} } \left[ 
 {1\over 2 a^2}\right.
 \nonumber \\
&&\qquad \left. +{1\over 2(a^2 \sin^2 \theta + b^2 \cos^2 \theta)}
 \right].
 \nonumber \\
 \end{eqnarray}
The bending contribution to the energy is then written as 
\begin{eqnarray}
F_\mathrm{bend}^\mathrm{ellips}(h) &=& {\pi k a^3 \over b^3} \int_{-1}^{y(h)} { \di y \over \sqrt{1-e^2 y^2} } \left[  {1\over 1- e^2 y^2} + {b^2 \over a^2}\right]^2
\nonumber \\
&=& \pi k \Big[ {2a y(h) \over b \sqrt{1-e^2 y(h)^2}} + {a^3 \over b^3} {3 y(h) -2 e^2 y(h)^3 \over 3  (1-e^2 y(h)^2)^{3/2}}
\nonumber \\
&& +{b\over a e} \left( \arcsin(e y(h)) +\arcsin e \right) +  { 2 a \over b \sqrt{ 1-e^2 } }
\nonumber \\
&& + {a^3 \over b^3} { 3 - 2 e^2 \over 3  ( 1 - e^2 )^{3/2}}
\Big],
\\
{\di F_\mathrm{bend}^\mathrm{ellips}(h) \over \di h} &=& { \pi k a^3/b^3 \over \sqrt{1-e^2 y(h)^2} } \left[ {1\over 1- e^2 y(h)^2} + {b^2\over a^2}\right]^2.
\nonumber
\end{eqnarray}
For spherical invaders the equations reduce to
\begin{eqnarray}
F_\mathrm{bend}^\mathrm{sph}(h)
= 4\pi {h  \kappa \over a} 
&\qquad & 
{\mathrm{d} F_\mathrm{bend}^\mathrm{sph}(h) \over \mathrm{d} h} = 4 \pi {\kappa \over a} 
\end{eqnarray}
\\
{\bf Line tension.}~This contribution describes the energy associated to the deformation of the membrane at the junction (triple line) between the contact region and the non-adhering surface of the host cell. As such 
\begin{eqnarray}
F^\mathrm{ellips}_\mathrm{lt}(h) &=& F^\mathrm{sph}_\mathrm{lt}(h) =  2\pi \gamma \Gamma(h)= 2\pi \gamma a \sqrt {1 - y(h)^2}
\nonumber \\
{\mathrm{d} F^\mathrm{ellips}_\mathrm{lt}(h) \over \mathrm{d} h} &=&  - {2\pi \gamma a \over b} { y(h) \over \sqrt {1 - y(h)^2} },
\end{eqnarray}
where $\gamma$ is the line tension and $\Gamma(h)$ is the length of the triple line.

\section{Calculation of the adhesion free energy for mobile ligands}\label{app:adhesion}

 We consider an invader of area $S_\mathrm{Tot}$ carrying $N_\mathrm{L}$ ligands, interacting with a cell surface functionalized by receptors. $S_\mathrm{CR}$ denotes the area of the contact region (CR) between the cell and the invader, while $S_\mathrm{OR}$ is the area of the invader outer region (OR) (see Fig.~\ref{Fig1}). The density of receptors in the CR is controlled by the areal density $\rho^{(0)}_\mathrm{R}$ of an ideal receptor reservoir in contact with the CR, related to a receptor chemical potential $\mu_\mathrm{R}$ by the relation $ \mu_\mathrm{R} \sim \log \rho^{(0)}_\mathrm{R}$.
 Ligands and receptors are modeled as freely diffusing hard disks.  
 Ligands can reversibly  bind receptors forming connections between the the invader and the cell membrane (see  Fig.~\ref{Fig1}). Reaction dynamics is controlled by the equilibrium constant $K^{(\mathrm{eq})}_\mathrm{2D}$ (see main text).\\
The partition function of the system ($Z$) is derived summing over all the possible configurations of the system, specified by the number of dimers ($n_D$), of receptors ($n_R$, $n_\mathrm{R}-n_D$ of which unbound), and ligands ($n_L$, $n_\mathrm{L}-n_D$ of which unbound) present in the CR 
\begin{widetext}
\begin{eqnarray}
Z &=& \sum_{n_\mathrm{L} = 0}^{N_\mathrm{L}} \sum_{n_\mathrm{R} \geq 0} \sum_{n_\mathrm{D}=0}^{\mathrm{min}[n_\mathrm{R},n_\mathrm{L}]} {\cal Z}(n_\mathrm{L},n_\mathrm{R},n_\mathrm{D})=\sum_{n_\mathrm{L} = 0}^{N_\mathrm{L}} \sum_{n_\mathrm{R} \geq 0} \sum_{n_\mathrm{D}=0}^{\mathrm{min}[n_\mathrm{R},n_\mathrm{L}]} \exp[-\beta  {\cal F}(n_\mathrm{L},n_\mathrm{R},n_\mathrm{D}) ]
\nonumber \\
{\cal Z}(n_\mathrm{L},n_\mathrm{R},n_\mathrm{D}) 
 &=&{N_\mathrm{L }\choose n_\mathrm{L} }  (S_\mathrm{OR})^{N_\mathrm{L }- n_\mathrm{L}} (S_\mathrm{CR})^{n_\mathrm{R}+n_\mathrm{L}-n_\mathrm{D}} Z_\mathrm{OR}^\mathrm{(excl)} (N_\mathrm{L} - n_\mathrm{L}) \times  
\label{eq:Z}
\\
&&   {(\rho^{(0)}_\mathrm{R})^{n_\mathrm{R}} \over n_\mathrm{R}!}  
  {n_\mathrm{R}! n_\mathrm{L}!   \over n_\mathrm{D}! (n_\mathrm{R}-n_\mathrm{D})!(n_\mathrm{L}-n_\mathrm{D})!}
 (K^\mathrm{(eq)}_\mathrm{2D})^{n_\mathrm{D}} Z_\mathrm{CR}^\mathrm{(excl)} (n_\mathrm{R}-n_\mathrm{D},n_\mathrm{L}-n_\mathrm{D},n_\mathrm{D}).
\nonumber
\end{eqnarray}
\end{widetext}
In Eq.~\ref{eq:Z}, $(\rho^{(0)}_\mathrm{R})^{n_\mathrm{R}} / n_\mathrm{R}!$ is the grand-canonical weight of having $n_\mathrm{R}$ receptors in the CR while the following combinatorial term accounts for the number of ways $n_\mathrm{D}$ dimers can be formed starting from $n_\mathrm{L}$ ligands and $n_\mathrm{R}$ receptors in the contact region \cite{francisco-pnas,shimobayashi2015direct}. ${\cal F}$ is the free-energy at fixed number of complexes in the CR, and 
$Z_\mathrm{OR}^\mathrm{(excl)}$ is the non-ideal part of the partition function of $N_\mathrm{L}-n_\mathrm{L}$ ligands confined in an area equal to $S_\mathrm{OR}$, and can be written as
\begin{eqnarray}
Z_\mathrm{OR}^\mathrm{(excl)} &=& {1\over (S_\mathrm{OR})^{N_\mathrm{L}-n_\mathrm{L}}} \int \mathrm{d}^2 {\bf r}_1 \cdots  \mathrm{d}^2 {\bf r}_{N_\mathrm{L}-n_\mathrm{L}} \cdot 
\label{eq:Zor}
\\
&& \qquad \qquad \cdot \exp[-\beta \sum_{i<j} V_\mathrm{LL}(|{\bf r}_i-{\bf r}_j  |)],
\nonumber
\end{eqnarray}
where ${\bf r}_i$ are ligand coordinates spanning the outer region of the invader, and $V_\mathrm{LL}$ models excluded volume interactions between ligands. 
We neglect curvature effects and calculate $Z_\mathrm{OR}^\mathrm{(excl)}$ using flat surfaces with periodic boundary conditions. This approximation is valid in the limit of big invaders and allows sampling the non-ideal properties of the system using small simulation boxes at given ligand and receptor densities.
$Z_\mathrm{CR}^\mathrm{(excl)} $ is the non-ideal part of the partition function in the contact region and is defined similarly to Eq.\ \ref{eq:Zor}. However $Z_\mathrm{CR}^\mathrm{(excl)}$ also includes excluded volume interactions between dimers and ligands/receptors as specified by the potentials $V_\mathrm{LL}$, $V_\mathrm{LD}$ and $V_\mathrm{RD}$. Without loss of generality in this study we have neglected ligand-receptor steric interactions. If one chose $V_\mathrm{LR}\neq 0$, however, the thermodynamic integration procedure defined in Eq.\ 5 of the main text should have included an extra contribution due to the fact that ligands and receptors interact also in absence of dimerization. 
In this work we sampled micro--states distributed as in Eq.\ \ref{eq:Zor} using Monte Carlo simulations (Sec.\ \ref{app:adhesion:MC}) and a virial expansion as detailed in Sec.\ \ref{app:adhesion:virial}.

\subsection{Monte Carlo algorithm}\label{app:adhesion:MC}

The Monte Carlo moves we implemented are sketched in Fig.~\ref{Fig1} of the main text. The acceptance rules presented below satisfy detailed balance conditions calculated using Eq.~\ref{eq:Z}.\\
Ligands are moved between the CR and the OR by means of a {\bf semi-grand canonical move} that conserves the total number of ligands.  The flow  chart of the algorithm is the following:
\begin{itemize}
\item{With equal probability we decide whether to attempt a displacement from the CR to the OR or vice versa.}
\item{A ligand is randomly chosen from the CR (OR).}
\item{A new position for the ligand is randomly selected in the OR (CR).}
\item{We check whether the new position satisfies excluded volume constraints (\emph{i.e.} the ligand does not overlap with another ligand or a dimer).}
\item{If excluded volume constraints are satisfied the move is accepted with probability
\begin{eqnarray}
\mathrm{acc}_\mathrm{disp} &=& \mathrm{min} \left[ 1, {n_\mathrm{o} S_\mathrm{n} \over (N_\mathrm{L}-n_\mathrm{o}+1) S_\mathrm{o}}\right],
\end{eqnarray}
where $n_\mathrm{o}$ is the number of ligands in the region from which we attempt to remove a binder, and $S_\mathrm{o}/S_\mathrm{n}$ (o/n=CR or OR) is the area of the old/new region.\\
}
\end{itemize}

Receptors are exchanged between the CR and an ideal reservoir with areal density $\rho^{(0)}_R$ by means of a {\bf grand-canonical move}, implemened as follows:
\begin{itemize}
\item{With equal probability we decide whether to attempt an insertion or removal of a receptor from the CR.}
\item{If an insertion move is chosen we randomly select a position for the new receptor in the CR.}
\item{We check excluded volume constraints in the CR.}
\item{If excluded volume constraints are satisfied we accept the insertion move with probability
\begin{eqnarray}
\mathrm{acc}_\mathrm{ins} &=& \mathrm{min} \left[ 1, {\rho^{(0)}_\mathrm{R} S_\mathrm{CR} \over m+1} \right],
\end{eqnarray}
where $m$ is the number of receptors in the CR prior the move.}
\item{For removal moves we chose a random receptor to remove from the CR}
\item{Removal moves are accepted with probability
\begin{eqnarray}
\mathrm{acc}_\mathrm{rem} &=& \mathrm{min} \left[ 1, {m\over \rho^{(0)}_\mathrm{R} S_\mathrm{CR}} \right].
\end{eqnarray}}
\end{itemize}

{\bf Reaction moves} in which dimers are formed from a dissociated ligand-receptor pair in the CR, or an exsisting dimer is split into a ligand and a receptor, are implemented as follows:
 \begin{itemize}
\item{With equal probability we decide whether to form or break a dimer.}
\item{If a dimer formation is attempted, we randomly chose a ligand and a receptor from the CR.}
\item{A position for the newly formed dimer is chosen randomly in the CR.}
\item{Excluded volume constraints are checked.}
\item{If excluded volume constraints are satisfied, the dimerisation is accepted with probability
\begin{eqnarray}
\mathrm{acc}_\mathrm{bind} &=& \mathrm{min}\left[ 1, { n m \over d+1 } { K^\mathrm{(eq)}_\mathrm{2D} \over  S_\mathrm{CR} } \right],
\nonumber \\
\end{eqnarray}
where $n$, $m$, and $d$ are, respectively, the number of ligands, receptors and dimers in the contact area prior the move.}
\item{If a dimer breakup is attempted, we randomly chose a dimer in the CR.}
\item{New positions for the freed ligand and receptor are chosen randomly in the CR.}
\item{Excluded volume constraints are checked for the ligand and the receptor.}
\item{If excluded volume constrains are satisfied the breakup move is accepted with probability 
\begin{eqnarray}
\mathrm{acc}_\mathrm{unbind}&=& \mathrm{min}\left[ 1,  { d \over (n+1)(m+1) } {  S_\mathrm{CR} \over K^\mathrm{(eq)}_\mathrm{2D} } \right].
\nonumber \\
\end{eqnarray}
}
\end{itemize}

\subsection{Virial expansion of the adhesion free energy}\label{app:adhesion:virial}

We estimate the partition function of the system (Eq.\ \ref{eq:Z}) by using a second virial approximation of the excluded volume part of for the OR ($Z^\mathrm{(excl)}_\mathrm{OR}$) and the CR ($Z^\mathrm{(excl)}_\mathrm{CR}$) partition functions \cite{hill2013statistical}. 
In terms of Mayer factors ($f_\mathrm{LL}(r_{ij}) = \exp[V_\mathrm{LL}(|{\bf r}_i -{\bf r}_j|)]-1$)  $Z^\mathrm{(excl)}_\mathrm{OR}$ (Eq.\ \ref{eq:Zor}) can be written as 
\begin{eqnarray}
Z_\mathrm{OR}^\mathrm{(excl)} (N_\mathrm{L} - n_\mathrm{L}) &=& \int { \mathrm{d} {\bf r}_1 \cdots \mathrm{d} {\bf r}_{N_\mathrm{L}-n_\mathrm{L}} \over (S_\mathrm{CR})^{N_\mathrm{L}-n_\mathrm{L}}} \prod_{i<j} \left(
f_{ij}(r_{ij})+1
\right)
\nonumber\\
&=& 1 + {1\over S_\mathrm{CR}^2} \sum_{i<j} \int  \mathrm{d} {\bf r}_i \mathrm{d} {\bf r}_j f_{ij}(r_{ij}) + \cdots,
\nonumber \\
\end{eqnarray}
from which we obtain
\begin{eqnarray}
Z_\mathrm{OR}^\mathrm{(excl)} (m) &=& 
1-B_2 {m (m-1) \over  S_\mathrm{OR} }   + {\cdots},
\label{eq:vir_OR}
\end{eqnarray}
where $B_2=\pi \alpha^2/2$ is the second virial coefficient of disks with diameter $\alpha$. 
Note that corrections to the previous expressions are of the order of $\phi^2$. Similarly 
\begin{eqnarray}
& Z_\mathrm{CR}^\mathrm{(excl)} ( n^0_\mathrm{R},n^0_\mathrm{L},n_\mathrm{D}) = 
1-B_2 { n^0_\mathrm{L} (n^0_\mathrm{L}-1) \over  S_\mathrm{CR} }   
-B_2 { n^0_\mathrm{R} (n^0_\mathrm{R}-1) \over  S_\mathrm{CR} } &
\nonumber \\
&-B_2 { n_\mathrm{D} (n_\mathrm{D}-1) \over  S_\mathrm{CR} }
 - 2 B_2 { n^0_\mathrm{L}  n_\mathrm{D} \over  S_\mathrm{CR} }
- 2 B_2 { n^0_\mathrm{R} n_\mathrm{D} \over  S_\mathrm{CR} } &
\label{eq:vir_CR}
\end{eqnarray}
where $n^0_\mathrm{L}$ and $n^0_\mathrm{R}$ are the numbers of free ligands and receptors in the CR ($n^0_\mathrm{L}= n_\mathrm{L}-n_\mathrm{D}$ and $n^0_\mathrm{R}= n_\mathrm{R}-n_\mathrm{D}$).
By inserting Eqs.\ \ref{eq:vir_OR} and \ref{eq:vir_CR} into Eq.\ \ref{eq:Z} we can explicitly estimate the free energy ${\cal F}$ as function of $n_\mathrm{L}$, $n_\mathrm{D}$,  $n_\mathrm{R}$, and $B_2$. In the saddle-point approximation we identify most probable values for the average numbers of ligands, receptors and dimers  by setting  
\begin{eqnarray}
{\mathrm{d}{\cal F} \over \mathrm{d}n_\mathrm{L} } = 0,   \qquad {\mathrm{d}{\cal F}\over \mathrm{d}n_\mathrm{R}} =0,    \qquad  {\mathrm{d}{\cal F} \over \mathrm{d}n_\mathrm{D} }=0  
\label{eq:SP}
\end{eqnarray}
obtaining, respectively 

\begin{eqnarray}
& \log \left[ {N_\mathrm{L} - n_L \over S_\mathrm{OR}} {S_\mathrm{CR} \over n_\mathrm{L}-n_\mathrm{D}} \right] = 2 B_2 \left[ {n_\mathrm{L}  \over S_\mathrm{CR}} - {N_\mathrm{L} - n_\mathrm{L} \over S_\mathrm{OR}}\right] &
\label{eq:saddle-L}
\\
& \log\left[ \rho^{(0)}_R S_\mathrm{CR} \over n_\mathrm{R}- n_\mathrm{D} \right] =   2 B_2  {n_\mathrm{R}\over S_\mathrm{CR}} &
\label{eq:saddle-R}
\\
& \log \left[ {(n_\mathrm{R}-n_\mathrm{D})(n_\mathrm{L}-n_\mathrm{D})\over n_\mathrm{D}} { K^\mathrm{(eq)}_\mathrm{2D} \over S_\mathrm{CR} } \right]  = - 2 B_2 { n_\mathrm{D} \over S_\mathrm{CR} }. &
\label{eq:saddle-D}
\end{eqnarray}
By setting $B_2=0$ in Eqs.\ \ref{eq:saddle-L}, \ref{eq:saddle-R} and \ref{eq:saddle-D} we obtain the number of ligands, receptors, and dimers ($n_{\mathrm{L},0}$, $n_{\mathrm{D},0}$, and $n_{\mathrm{R},0}$) in the limit of ideal linkers ($\alpha=0$)~\cite{shimobayashi2015direct} 
\begin{eqnarray}
n_{\mathrm{L},0} &=& N_\mathrm{L}  
{ S_\mathrm{CR}(1+K^\mathrm{(eq)}_\mathrm{2D} \rho^{(0)}_\mathrm{R} ) \over S_\mathrm{OR} + S_\mathrm{CR}(1+K^\mathrm{(eq)}_\mathrm{2D} \rho^{(0)}_\mathrm{R} ) }
\label{eq:ideal-L}
\\
n_{\mathrm{D},0} &=& N_\mathrm{L}  
{ S_\mathrm{CR} K^\mathrm{(eq)}_{2D} \rho^{(0)}_\mathrm{R}  \over S_\mathrm{OR} + S_\mathrm{CR}(1+K^\mathrm{(eq)}_\mathrm{2D} \rho^{(0)}_\mathrm{R} ) }
\label{eq:ideal-D}
\\
n_{\mathrm{R},0} &=& \rho^{(0)}_\mathrm{R} S_\mathrm{CR} + n_{\mathrm{D},0}
\label{eq:ideal-R}
\end{eqnarray}
Using Eqs. \ref{eq:ideal-L}, \ref{eq:ideal-D}, and \ref{eq:ideal-R} with Eqs.\ \ref{eq:saddle-L}, \ref{eq:saddle-R}, and \ref{eq:saddle-D} 
 we can calculate the leading order corrections to the ideal terms ($n_{\mathrm{L},1}$, $n_{\mathrm{D},1}$, and $n_{\mathrm{R},1}$). These satisfy 
\begin{eqnarray}
 & n_{\mathrm{D},1} - n_{\mathrm{L},1} \left[ {S_\mathrm{CR}\over S_\mathrm{OR}} + 1 \right] = { 2 B_2 n_{\mathrm{D},0} \over S_\mathrm{CR} } (n_{\mathrm{L},0} - n_{\mathrm{D},0})  &
\label{eq:nid-L}
\\
& n_{\mathrm{D},1}-n_{\mathrm{R},1} = 2 \rho^{(0)}_\mathrm{R}  B_2  n_{\mathrm{R},0} &
\label{eq:nid-R}
\\
& {n_{\mathrm{R},1}- n_{\mathrm{D},1} \over \rho^{(0)}_\mathrm{R} S_\mathrm{CR} } +{ n_{\mathrm{L},1} - n_{\mathrm{D},1} \over n_{\mathrm{L},0}-n_{\mathrm{D},0}} - { n_{\mathrm{D},1} \over n_{\mathrm{D},0} }  = - 2 B_2 { n_{\mathrm{D},0} \over S_\mathrm{CR} } &
\label{eq:nid-D}
\end{eqnarray}
The free energy of the system can then be calculated using the equilibrium concentrations of the complexes (Eqs.\ \ref{eq:ideal-L}-\ref{eq:nid-D}) in the perturbative expansion of ${\cal F}$  (Eqs.\ \ref{eq:Z}, \ref{eq:vir_OR}, and \ref{eq:vir_CR}) 
\begin{eqnarray}
\beta F&=& 
\beta {\cal F}(n_{\mathrm{L},0}+n_{\mathrm{L},1},n_{\mathrm{R},0}+n_{\mathrm{R},1},n_{\mathrm{D},0}+n_{\mathrm{D},1}) 
\nonumber \\
&\approx&  \beta {\cal F}(n_{\mathrm{L},0},n_{\mathrm{R},0},n_{\mathrm{D},0})
\nonumber\\
&=&K+N_\mathrm{L} \log {   N_\mathrm{L} - n_{\mathrm{L},0}  \over S_\mathrm{OR} }  
\nonumber \\
&& + {B_2 \over  S_\mathrm{CR}} \Big[ 
(n_{\mathrm{L},0} - n_{\mathrm{D},0} )( N_\mathrm{L}+n_{\mathrm{D},0} ) + n_{\mathrm{R},0}^2
\Big]
\nonumber \\
F^{\mathrm{mob},\alpha}_\mathrm{adh} &=& F(K^\mathrm{(eq)}_\mathrm{2D}) - F_0
\label{eq:free-en-per}
\end{eqnarray}
where $K=-N_\mathrm{L}\log N_\mathrm{L}-\rho^{(0)}_\mathrm{R} S_\mathrm{CR}$, $F_0$ is the reference value of the free-energy  that is calculated using $K^\mathrm{(eq)}_\mathrm{2D}=0$, and $S_\mathrm{Tot}$ is the total area of the invader ($S_\mathrm{Tot}=S_\mathrm{OR}+S_\mathrm{CR}$). Note that because of the saddle-point equations \ref{eq:SP}, at the leading order in $\phi$, only the ideal number densities contribute to the free energy. Also, $F_0$ should be subtracted from $F$ when calculating adhesion free energies.Using Eqs.\ \ref{eq:ideal-R}, \ref{eq:ideal-L}, \ref{eq:ideal-D} in Eq.\ \ref{eq:free-en-per} we can derive Eq.~\ref{eq:ideal} and \ref{eq:excess}.

\section{ Calculation of the adhesion free energy for fixed ligands}\label{app:adhesion:fixed}

Here we adapt calculations of the previous section to the case of invaders decorated by fixed ligands binding ideal mobile receptors. In this case the number of ligands in the CR is fixed and equal to $n_\mathrm{L}(h)=N_\mathrm{L} S_\mathrm{CR}(h)/S_\mathrm{Tot}$. Similarly to Eq.\ \ref{eq:Z} the partition function is then given by 
\begin{eqnarray}
&Z = \sum_{n_\mathrm{R} \geq 0} \sum_{n_\mathrm{D}=0}^{\mathrm{min}[n_\mathrm{R},n_\mathrm{L}]} {\cal Z}(n_\mathrm{R},n_\mathrm{D})  &
\label{eq:Zfixed} \\
& {\cal Z}(n_\mathrm{R},n_\mathrm{D}) 
=   {(S_\mathrm{CR} \rho^{(0)}_\mathrm{R})^{n_\mathrm{R}} \over n_\mathrm{R}!}  
{ n_\mathrm{R} \choose n_\mathrm{D} } { n_\mathrm{L}(h) \choose n_\mathrm{D} } n_\mathrm{D}! \left( K^\mathrm{(eq)}_\mathrm{2D} \over S_\mathrm{CR} \right)^{n_\mathrm{D}} \, . &
\nonumber
\end{eqnarray}
Using the previous equations we can calculate the average number of ideal receptors and dimers by solving the saddle--point equations 
\begin{eqnarray}
{\mathrm{d} {\cal F}(n_{\mathrm{R},0}, n_{\mathrm{D},0}) \over \mathrm{d} n_\mathrm{R}} = 0 \qquad  {\mathrm{d} {\cal F}(n_{\mathrm{R},0}, n_{\mathrm{D},0}) \over \mathrm{d} n_\mathrm{D}} = 0,
\end{eqnarray}
obtaining 
\begin{eqnarray}
&n_{\mathrm{R},0} - n_{\mathrm{D},0} =  \rho^{(0)}_\mathrm{R} S_\mathrm{CR} &  
\nonumber \\
& {n_{\mathrm{D},0} \over (n_{\mathrm{R},0} - n_{\mathrm{D},0})(n_{\mathrm{L},0} - n_{\mathrm{D},0}) } = { K^\mathrm{(eq)}_\mathrm{2D} \over S_\mathrm{CR} }.&\\ \nonumber
\label{eq:SPfixed}
\end{eqnarray}

Using Eqs.\ \ref{eq:SPfixed} into Eq.\ \ref{eq:Zfixed} we can calculate the free energy and adhesion free energy as follows
\begin{eqnarray}
\beta F &=&  \beta {\cal F}(n_{\mathrm{R},0},n_{\mathrm{L},0})  
\nonumber \\
&=& -n_\mathrm{L}(h) \log (1 + K^\mathrm{(eq)}_\mathrm{2D} \rho^{(0)}_\mathrm{R})  + \rho^{(0)}_\mathrm{R} S_\mathrm{CR} 
\nonumber
\\
F^{\mathrm{fix},\alpha=0}_\mathrm{adh} &=& F(K^\mathrm{(eq)}_\mathrm{2D}) - F(0)
\label{eq:adhfix}
\end{eqnarray} 
Eq.\ \ref{eq:adhfix} corresponds to Eq.~\ref{eq:fixed}.

}



\providecommand{\noopsort}[1]{}\providecommand{\singleletter}[1]{#1}%
%


\end{document}